\documentclass[%
 aip,
  rsi,
 amsmath,amssymb,
 preprint,review, 12pt
]{revtex4-1}

\usepackage{graphicx}
\usepackage{dcolumn}
\usepackage{bm}

\usepackage[utf8]{inputenc}
\usepackage[T1]{fontenc}
\usepackage{mathptmx}
\usepackage{IEEEtrantools}

\begin{document}

\preprint{Draft}

\title{\large \centering Characterization of Commercial Thermoelectric Modules for Precision Heat Flux Measurement}

\author{Jacob Crossley}
 \affiliation{Department of Mechanical Engineering, University of Utah, Salt Lake City, Utah 84112, USA}%

\author{A. N. M. Taufiq Elahi }%
 \affiliation{Department of Mechanical Engineering, University of Utah, Salt Lake City, Utah 84112, USA}%

\author{Mohammad Ghashami}
\email{mghashami2@unl.edu}
\affiliation{Mechanical \& Materials Engineering Department, University of Nebraska-Lincoln, Lincoln, Nebraska 68588, USA}

\author{Keunhan Park}
\email{kpark@mech.utah.edu}
\affiliation{Department of Mechanical Engineering, University of Utah, Salt Lake City, Utah 84112, USA}%

\date{\today}

\begin{abstract}
In this article, we present a cost-effective approach to the precision measurement of heat flux using commercial thermoelectric modules (TEMs). Two different methods of measuring heat flux with TEMs are investigated, namely passive mode based on the Seebeck effect and active mode based on the Peltier effect.
For both modes, a TEM as a heat fluxmeter is calibrated to show a linear relation between the voltage across the TEM and the heat flow rate from 0 to 100 mW. While both modes exhibit sufficiently high sensitivities suitable for low heat flow measurement, active mode is shown to be $\sim$7 times more sensitive than passive mode. From the speculation on the origin of the measurement uncertainty, we propose a dual TEM scheme by operating the top TEM in passive mode while its bottom temperature maintains constant by the feedback-controlled bottom TEM. The dual-TEM scheme can suppress the sensitivity uncertainty by up to 4 times when compared to the single-TEM passive mode by stabilizing the bottom temperature. The response time of a 1.5 cm $\times$ 1.5 cm TEM is measured to be $8.90 \pm 0.97$ seconds for heating and $10.83  \pm 0.65$ seconds for cooling, which is slower than commercial heat fluxmeters but still fast enough to measure heat flux with a time resolution on the order of 10 seconds. We believe that the obtained results can facilitate the use of a commercial TEM for heat flux measurement in various thermal experiments. 
\end{abstract}

\maketitle

\section{Introduction}
Measurement of heat flux has been a crucial part of thermal metrology, as knowledge of heat flux can be used to verify theory or provide meaningful information in engineering design. Thermoelectric heat flux sensors have been used to measure heat flux in conduction, convection, and radiation systems to assist in developing more accurate modeling of their respective systems.\cite{Wong1979,Cherif2009,Santos2014,Kobari2015} However, most of commercially available heat flux sensors are expensive and typically designed for high heat flux measurement, which is not suitable for low power or microscopic systems. 
In recent years, thermoelectric modules (TEMs) have seen more use as a low-cost alternative for measuring heat flux.\cite{Haruyama2001,Mann2006,Sippawit2015,Song2020,ChenAustin2021,Ahamat2017,Rizzo2021} 
While TEMs are used in a wide variety of applications, they are most commonly used as either a cooler or a power generator.\cite{Min2006,He2015} Thermoelectric coolers function as a solid state heat pump by receiving electrical power and uses the cold side of the TEM to remove heat from a system. On the other hand, a thermoelectric generator work as a solid state heat engine and takes advantage of the Seebeck effect to generate electrical power by means of a temperature difference. It should be noted that commercially available thermoelectric heat fluxmeters, including the previous attempts to use a TEM for heat flux measurement, are based on the Seebeck effect; although, the measured electrical voltage is used to determine the heat flux instead of generating power \cite{Haruyama2001,Sippawit2015,Ahamat2017,ChenAustin2021,Rizzo2021}. 

Although a TEM and a conventional heat fluxmeter operate on the same principle, the TEM sacrifices a size and a faster response time to achieve a higher voltage signal ranging from $\mathrm{\mu V}$ to $\mathrm{mV}$. For a steady state or slow response system, a TEM can prove more useful in measuring heat flux over a conventional sensor.\cite{Ahamat2017,ChenAustin2021}
Here we discuss and demonstrate the use of a commercial TEM as a heat fluxmeter with two operating schemes, namely passive and active modes. Passive mode is essentially the same as a conventional themoelectric heat flux measurement based on the Seebeck effect, which measures an output voltage induced by the temperature difference between the top and bottom junctions.  
On the other hand, the active mode operation is based on the Peltier effect by flowing an electric current through the TEM to induce a temperature difference between the top and bottom junctions.\cite{Vidhya2018} While the TEM essentially works as a cooler in active mode to remove heat from the heat source, heat flux to the TEM can be determined by the supplied voltage to maintain the top junction temperature at a set point. In the following section, we discuss the theoretical background of using a TEM in passive and active modes for heat flux measurement. TEM devices are then systematically calibrated for passive and active modes under steady and transient states. Based on the calibration results for passive-mode and active-mode heat flux measurements, we propose dual-stack mode as a new heat flux measurement scheme that can improve the precision up to four times when compared to passive mode by maintaining the bottom temperature of the TEM heat fluxmeter with another feedback-controlled TEM. This result suggests that dual-stack mode should be adopted as a standard heat flux measurement scheme when a commercial TEM is to be used as a precision heat fluxmeter. 

\section{Theoretical Background}
\begin{figure}[t!]
\includegraphics[width=\textwidth,height=\textheight,keepaspectratio]{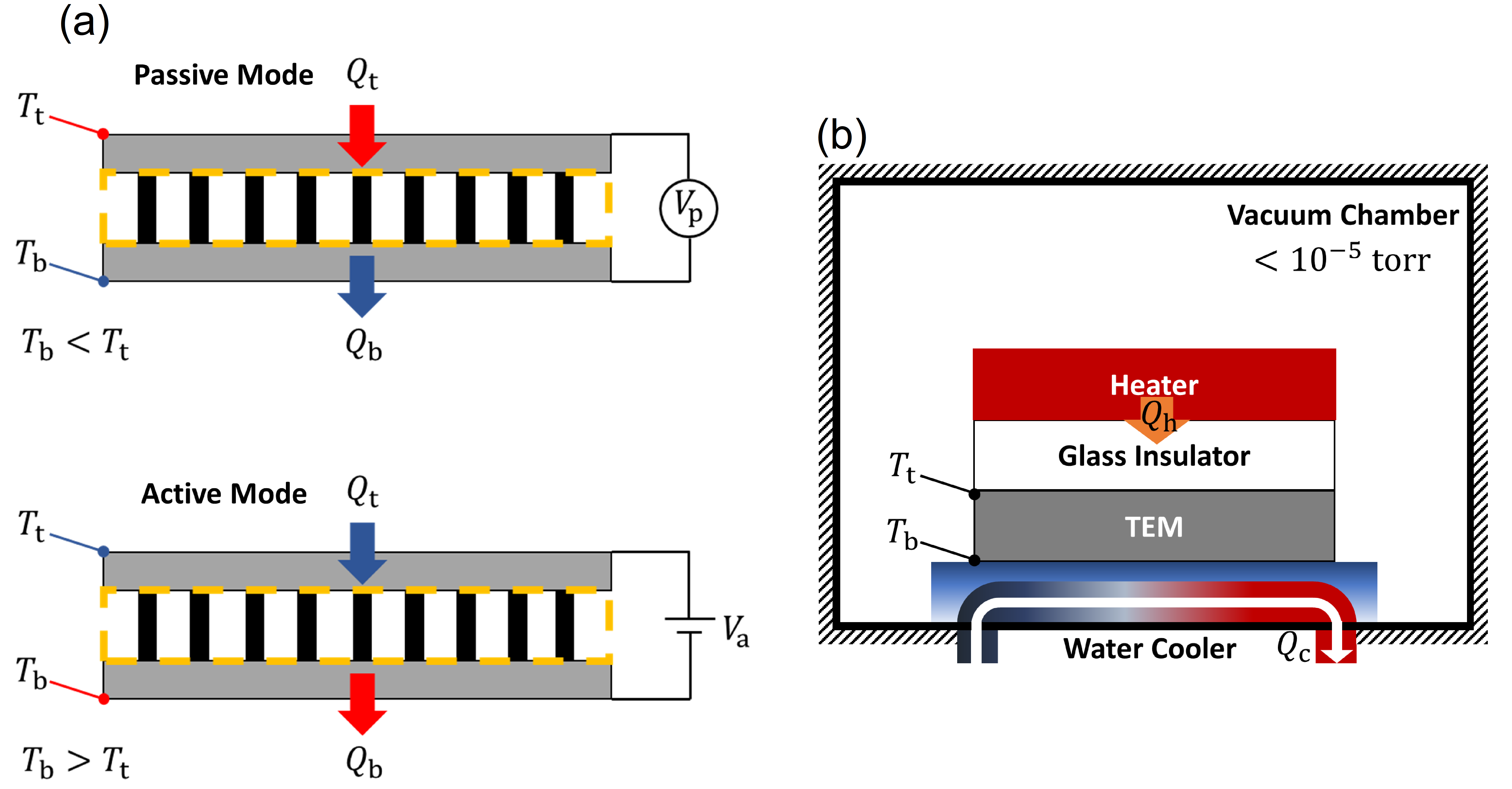}
\caption{Schematic illustration of the TEM setup in (a) passive mode and (b) active mode. In passive mode, $T_\mathrm{t} > T_\mathrm{b}$. However, in the active case, the opposite is true due to the Peltier effect caused by current flow.}
\label{Schematic}
\end{figure}

In this section, we show analytical relations between the heat flow rate through the TEM and its voltage signal for both passive and active modes: the schematic diagrams are shown in Fig. \ref{Schematic}(a). In both cases, heat is assumed to be transferred to the top of the TEM while the bottom temperature is maintained constant at $T_\mathrm{b}$. In passive mode, the heat flow rates through the TEM can be written as
\begin{equation}
 Q_{\mathrm{t}} =  Q_{\mathrm{b}} = \frac{T_\mathrm{t} - T_\mathrm{b}}{R_\mathrm{th}},
 \label{Eq:Fourier}
\end{equation}
where $Q_{\mathrm{t}}$ ($Q_{\mathrm{b}}$) is the heat flow rate through the top (bottom) plate of the TEM at temperature $T_\mathrm{t}$ ($T_\mathrm{b}$), and $R_\mathrm{th}$ is the thermal resistance of the TEM. Since there is no electric current applied to the device, $Q_{\mathrm{t}}$ should be the same as $Q_{\mathrm{b}}$ if there is no heat loss from the TEM. The voltage across the TEM is proportional to the temperature difference according to the Seebeck effect, i.e., $V_\mathrm{p} = \alpha_\mathrm{eff} (T_\mathrm{t} - T_\mathrm{b})$, where $\alpha_\mathrm{eff}$ is the effective Seebeck coefficient of the TEM. Therefore, the voltage output in passive mode $V_\mathrm{p}$ can be written in linear relation with $Q_\mathrm{t}$ as \cite{Ahamat2017,Rizzo2021} 
\begin{equation}
V_\mathrm{p}= \left(\alpha_\mathrm{eff} R_\mathrm{th}\right)Q_\mathrm{t} + V_\mathrm{p,0}.
 \label{Eq:Q_passive}
\end{equation}
Here, $V_\mathrm{p,0}$ is added to reflect the offset voltage output under the zero heater power due to the intrinsic thermal non-equilibrium existing in the experimental setup. From the above equation, the sensitivity of a TEM device in passive mode can be defined as $S_\mathrm{p}\equiv V_\mathrm{p}/Q_\mathrm{t}=\alpha_\mathrm{eff} R_\mathrm{th}$.

When the TEM is operated in active mode to maintain $T_\mathrm{t}$ as shown in the bottom of Fig. \ref{Schematic}(a), an energy balance analysis (neglecting Thompson effect) at the top and bottom junctions of the TEM yields the following equations: \cite{Lineykin2007,Zhang2014,Zhao2014,Ahamat2017}
\begin{subequations}
    \label{Eq:Q_active}
    \begin{IEEEeqnarray}{rCl}
    Q_{\mathrm{t}} &=  \alpha_\mathrm{eff} I T_\mathrm{t} - \frac{1}{2}I^2 R_e - \frac{T_\mathrm{b} - T_\mathrm{t}}{R_\mathrm{th}} \\
    \nonumber \\
   Q_{\mathrm{b}} &=  \alpha_\mathrm{eff} I T_\mathrm{b} + \frac{1}{2}I^2 R_e - \frac{T_\mathrm{b} - T_\mathrm{t}}{R_\mathrm{th}},
\end{IEEEeqnarray}
\end{subequations}
where $I$ is the electrical current supplied to the TEM having the electrical resistance $R_e$. It should be noted that the TEM is essentially operated as a cooler in active mode (i.e., $T_\mathrm{t} < T_\mathrm{b}$). The first terms of Eqs. (\ref{Eq:Q_active}a) and (\ref{Eq:Q_active}b) are the heat absorption and generation rates at the cold top and hot bottom junctions, respectively, while the second and the third terms denote heat conduction from the hot to the cold junction under the joule heating. 
The electrical power input to the TEM for active mode is thus written as $P_\mathrm{e}=Q_{\mathrm{b}} -  Q_{\mathrm{t}} = \alpha_\mathrm{eff} I (T_\mathrm{b} - T_\mathrm{t}) + I^2R_e$, from which the applied voltage in active mode is written as 
\begin{equation}
V_\mathrm{a} = P_\mathrm{e}/I = \alpha_\mathrm{eff} (T_\mathrm{b} - T_\mathrm{t}) + IR_e.
 \label{Eq:V_a}
\end{equation}
If the joule heating effect in Eq. (\ref{Eq:Q_active}a) is ignored due to low-current operations (i.e., $I \lessapprox 10 \ \mathrm{mA}$ for a 1.5 cm $\times$ 1.5 cm TEM device) for small heat flux measurement, the heat flow rate to the TEM can be approximated as a function of $V_\mathrm{a}$ to yield
\begin{equation}
Q_{\mathrm{t}} =  \left(\frac{\alpha_\mathrm{eff} T_\mathrm{t}}{R_e}\right)V_\mathrm{a} - \left(T_\mathrm{b} - T_\mathrm{t}\right)\left(\frac{\alpha_\mathrm{eff}^2 T_t}{R_e}+\frac{1}{{R_\mathrm{th}}}\right).
 \label{Eq:Q_t}
\end{equation}
When there is no heat flow to the TEM (i.e., $Q_{\mathrm{t}}=0$), the TEM voltage can be written as $V_\mathrm{a,0}=\left(T_\mathrm{b} - T_\mathrm{t}\right)\left[\alpha_\mathrm{eff} + R_e/\left(\alpha_\mathrm{eff} T_\mathrm{t} R_\mathrm{th}\right)\right]$, which is due to the electric current required to maintain the pre-set temperature difference between the top and bottom junctions of the TEM. Therefore, the voltage output in active mode $V_\mathrm{a}$  can be written in linear relation with $Q_\mathrm{t}$ as
\begin{equation}
V_\mathrm{a} = \left(\frac{R_e}{\alpha_\mathrm{eff} T_\mathrm{t}}\right)Q_\mathrm{t} + V_\mathrm{a,0}, 
 \label{Eq:Q_t_simple}
\end{equation}
from which the sensitivity of a TEM device for active mode can be written as $S_\mathrm{a}\equiv V_\mathrm{a}/Q_\mathrm{t}=R_\mathrm{e}/(\alpha_\mathrm{eff} T_\mathrm{t})$. In Eqs. (\ref{Eq:Q_passive}) and (\ref{Eq:Q_t_simple}), it is clear that a TEM device can be used as a heat flux meter as the voltage output is linearly proportional to the heat flow rate for both passive and active modes. 


\section{Experimental Methods}
As discussed in the previous section, a TEM device should be calibrated first before implementing it for heat flux measurement. Figure \ref{Schematic}(b) illustrates the schematic diagram of the calibration setup. The entire setup is placed in a vacuum chamber (at $\sim 10^{-6}$ torr) to preserve a 1-D conduction condition. It is assumed that all heat generated by the heater is conducted through the TEM, $Q_\mathrm{h} = Q_\mathrm{t}$ in Fig.\ref{Schematic}(a), by ignoring the radiative heat loss from the top surface of the heater and from the side walls of the calibration setup. For all experiments, a ceramic heater (Thorlabs HT24S, 20$\times$20 mm$^2$) is in-line with a TEM (Custom Thermoelectric 03111-5L31-04CF, 15$\times$15 mm$^2$) with a soda-lime glass insulator in between. Presence of the glass insulator slows down the thermal response of the calibration system to secure sufficient time to feedback-control the TEM in active mode operation for stable maintenance of $T_\mathrm{t}$. Although not shown in Fig.\ref{Schematic}(a) for brevity, a 2-mm thick copper plate is placed between the heater and the glass insulator to make a uniform heat flux distribution across the cross section of the TEM while suppressing the undesired radiative heat transfer from the slightly larger heater to the TEM. We estimate that the heat loss due to far-field radiation to the surroundings at the maximum heater power is less than 1$\%$ of the total power supplied to the heater. 

It should be noted that the temperature at the bottom of the TEM is not actively controlled during either passive and active mode operations and is thus subject to change. Instead, a water chiller (Isotemp 4100 R35) is used to remove heat from the TEM. The water chiller is operated throughout the entire experiment to maintain a water bath temperature at $25^\circ$ C, which allows $T_{\mathrm{b}}$ at $24.59 \pm 0.12^\circ \mathrm{C}$ when the heater is turned off. When the maximum heater power is applied to the heater, the average $T_{\mathrm{b}}$ changes by about $0.14^\circ$C for passive-mode and $0.18^\circ$C for active-mode operations, respectively. 
The bottom temperature increases more in active mode due to power input to the TEM for cooling. The change and uncertainty of $T_{\mathrm{b}}$ throughout the experiment is believed to be cause of the slope uncertainty as will be discussed in the Results and Discussion section.
During experiments, the top and bottom temperatures of the TEM are monitored by commercial resistive temperature detectors (RTDs, OMEGA F3105), which have a nominal temperature measurement uncertainty of $ \pm 0.03^\circ \mathrm{C}$. 
For passive mode, the temperature at the top surface of the TEM ($T_\mathrm{t}$) is allowed to change depending on the heat flow rate from the heater, which determines the voltage output from the TEM. For active mode, on the other hand, the voltage input to the TEM is feedback-controlled under the following proportional-integral-derivative (PID) scheme:\cite{Karl1995,Aidan2009,Jarzembski2018}
\begin{equation}
V_\mathrm{a}(t)=G_\mathrm{P}e(t) + G_\mathrm{I}\int_0^t e(\tau)d\tau + G_\mathrm{D}\frac{de}{dt},
 \label{Eq:PID}
\end{equation}
where $e(t)$ is the instantaneous tracking error computed based on the difference between the transient $T_\mathrm{t}(t)$ and the user-defined temperature setpoint $T_\mathrm{SP}$, and $G_\mathrm{P}$, $G_\mathrm{I}$, and $G_\mathrm{D}$ are the proportional, integral, and derivative gains, respectively. Under the PID control scheme, the real-time change of $V_\mathrm{a}$ is monitored while the top surface temperature of the TEM is maintained at $T_\mathrm{t} = 23.45^\circ \mathrm{C} \pm 0.09^\circ \mathrm{C}$, slightly lower than the bottom surface temperature ($T_{\mathrm{b}}\approx24.6^\circ\mathrm{C}$).

\section{Results and Discussion}
\subsection{Steady-State Device calibration}
\begin{figure*}[t!]
\includegraphics[width=\textwidth,height=\textheight,keepaspectratio]{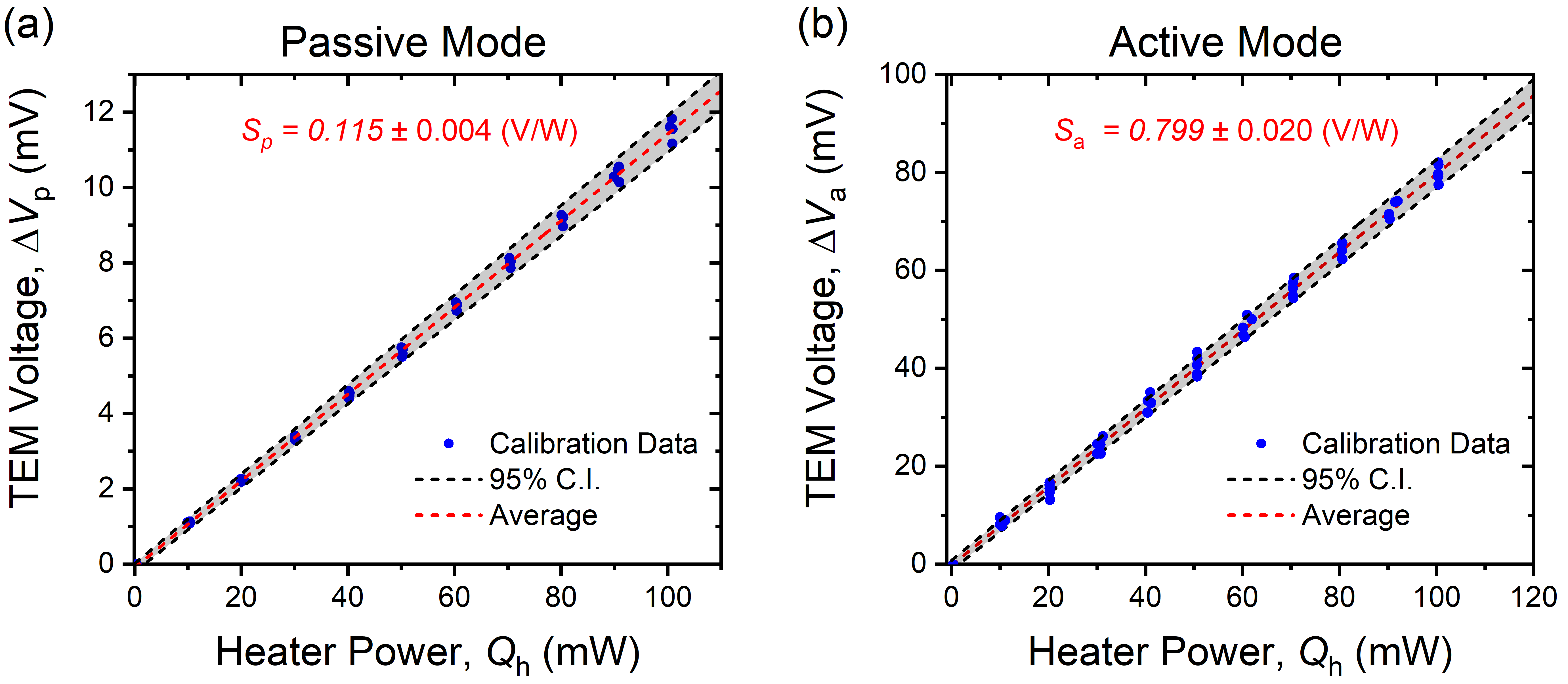}
\caption{Calibration data of the TEM in (a) passive mode and (b) active mode. The red dashed line is the average slope of each data set while the black dashed lines indicate a 95\% projected confidence interval. The sensitivity uncertainty is given as $\pm 0.004$ V/W for passive mode and $\pm 0.020$ W/V for active mode, which corresponds to $\pm 3.5 \%$ and $\pm 2.5 \%$}
\label{calibration}
\end{figure*}

Fig. \ref{calibration} shows the steady-state calibration results for (a) passive and (b) active modes for the heater power up to 100 mW with a 10-mW increment. For calibration, data points are collected to obtain the TEM voltage upon reaching a steady state at each heater power. In both cases, there are linear relations between the measured/applied TEM voltage and the applied heater power. The TEM sensitivity is measured to be $S_\mathrm{p}=0.115 \pm 0.004$ V/W for passive mode and $S_\mathrm{a}=0.799 \pm 0.020$ V/W for active mode, respectively. Here, the sensitivity uncertainty is determined from obtained data sets (i.e., 4 and 5 sets for passive and active mode, respectively) within a 95\% confidence interval, which are shown with the shaded region in the figure. The sensitivity of active mode is approximately 7 times greater than that of passive mode while exhibiting similar relative uncertainties ($\pm 3.5\%$ in passive mode and $\pm 2.5\%$ for active mode). This high sensitivity in active mode is beneficial particularly for low heat flux measurement. The noise spectral density of the instrument used for TEM calibration is measured to be around 10 $\mathrm{nV/\sqrt{Hz}},$\cite{Hamian2016} which corresponds to the noise equivalent heat flow rate (NEQ) of $\sim 86$ $\mathrm{nW/\sqrt{Hz}}$ for passive mode and $\sim 13$ $\mathrm{nW/\sqrt{Hz}}$ for active mode. However, it should be noted that active mode requires more sophisticated instrumentation for feedback control while passive mode can be easily implemented for heat flux measurement. An offset voltage under no heat flux ($V_\mathrm{a,0}$) should also be carefully calibrated for the accurate heat flux measurement under active mode.

The experimentally obtained sensitivities from the slopes of Fig. \ref{calibration} can be compared with the theoretically derived sensitivities, $S_\mathrm{p}=\alpha_\mathrm{eff} R_\mathrm{th}$ and $S_\mathrm{a}=R_\mathrm{e}/(\alpha_\mathrm{eff} T_\mathrm{t})$. As described in the previous section, the top and bottom temperatures of the TEM ($T_\mathrm{t}$ and $T_\mathrm{b}$) are measured along with the voltage across the TEM. From the passive-mode measurement, the effective Seebeck coefficient of the TEM is determined by $\alpha_\mathrm{eff}=V_\mathrm{p}/(T_\mathrm{t} - T_\mathrm{b})$. Similarly, the thermal resistance of the TEM ($R_\mathrm{th}$) can be determined by the temperature difference and the heater power from Eq. (\ref{Eq:Fourier}). The effective Seebeck coefficient and the thermal resistance of the TEM used in the present work are measured to be $\alpha_\mathrm{eff} = 13.076 \pm 0.620 \ \mathrm{mV/K}$ and $R_\mathrm{th} = 8.875 \pm 0.792 \ \mathrm{K/W}$, respectively, from which the sensitivity of the passive-mode operation is determined as $S_\mathrm{p}=0.116 \pm 0.016$  V/W. In addition, the electrical resistance of the TEM used for the present work is measured to be $R_\mathrm{e}=3.292 \pm 0.112 \ \Omega$. Therefore, the sensitivity of the active-mode operation can be calculated to be $S_\mathrm{a}=0.849 \pm 0.069$  V/W when the top junction temperature of the TEM is set to $T_\mathrm{t} = 23.45^\circ \mathrm{C}$. The calculated passive- and active-mode sensitivity values are in good agreement with the measurements, which validates the theoretical model of the present work.   

\subsection{Dynamic Response Characterization}

\begin{figure*}[b!]
\includegraphics[width=\textwidth,height=\textheight,keepaspectratio]{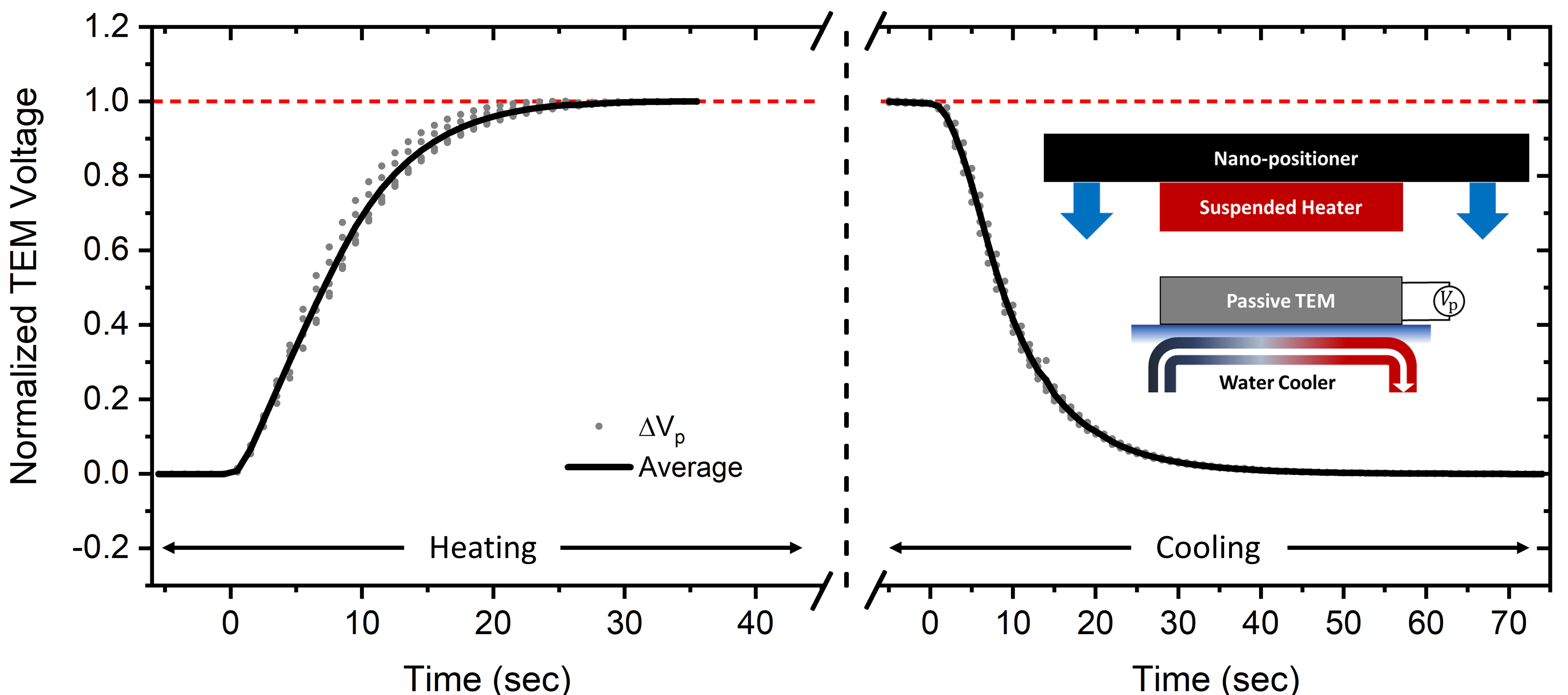}
\caption{The dynamic voltage response of the TEM obtained by suspending a heat source and bringing it into contact with the TEM. The data is normalized and the black curve indicates the average measured TEM voltage signal, while the gray dots indicate individual voltage measurements for four separate measurements. The TEM voltage reaches a steady state in less than a minute, revealing its thermal time constant to be $8.90 \pm 0.97$ sec for heating and $10.83  \pm 0.65$ sec for cooling.}
\label{dynamic response}
\end{figure*}

The response time of a heat fluxmeter is crucial for the accurate measurement of transient heat transfer phenomena. In the present work, the dynamic thermal response of the TEM is characterized by measuring the transient TEM voltage when a heater is made in contact with and retracted from the TEM. To this end, a heater is mounted on a nanopositioner (SMARPOD 110.45), which is housed in a vacuum chamber in a upside-down manner to place the heater slightly above the TEM on the fixed bottom stage: see the inset of Fig. \ref{dynamic response} for schematic illustration. The nanopositioner can precisely control the position of the heater with respect to the TEM. The heater is then powered to 100 mW, and brought into contact with the TEM device for heating or retracted from it for cooling while the TEM voltage is measured.  Figure \ref{dynamic response} shows the measured dynamic response of the TEM for both heating and cooling. An average time constant is extracted from 5 data sets, yielding $\tau_\mathrm{heat} = 8.90 \pm 0.97$ sec for heating and $\tau_\mathrm{cool} = 10.83  \pm 0.65$ sec for cooling. Here, the time constant is defined as the amount of time that the first order response takes to reach $63.2 \%$ of the steady state value. It should be noted that the obtained time constant is specific for the used TEM, and a faster response would be expected if a smaller TEM was used for heat flux measurement. 

\begin{figure*}[t!]
\includegraphics[width=\textwidth,height=\textheight,keepaspectratio]{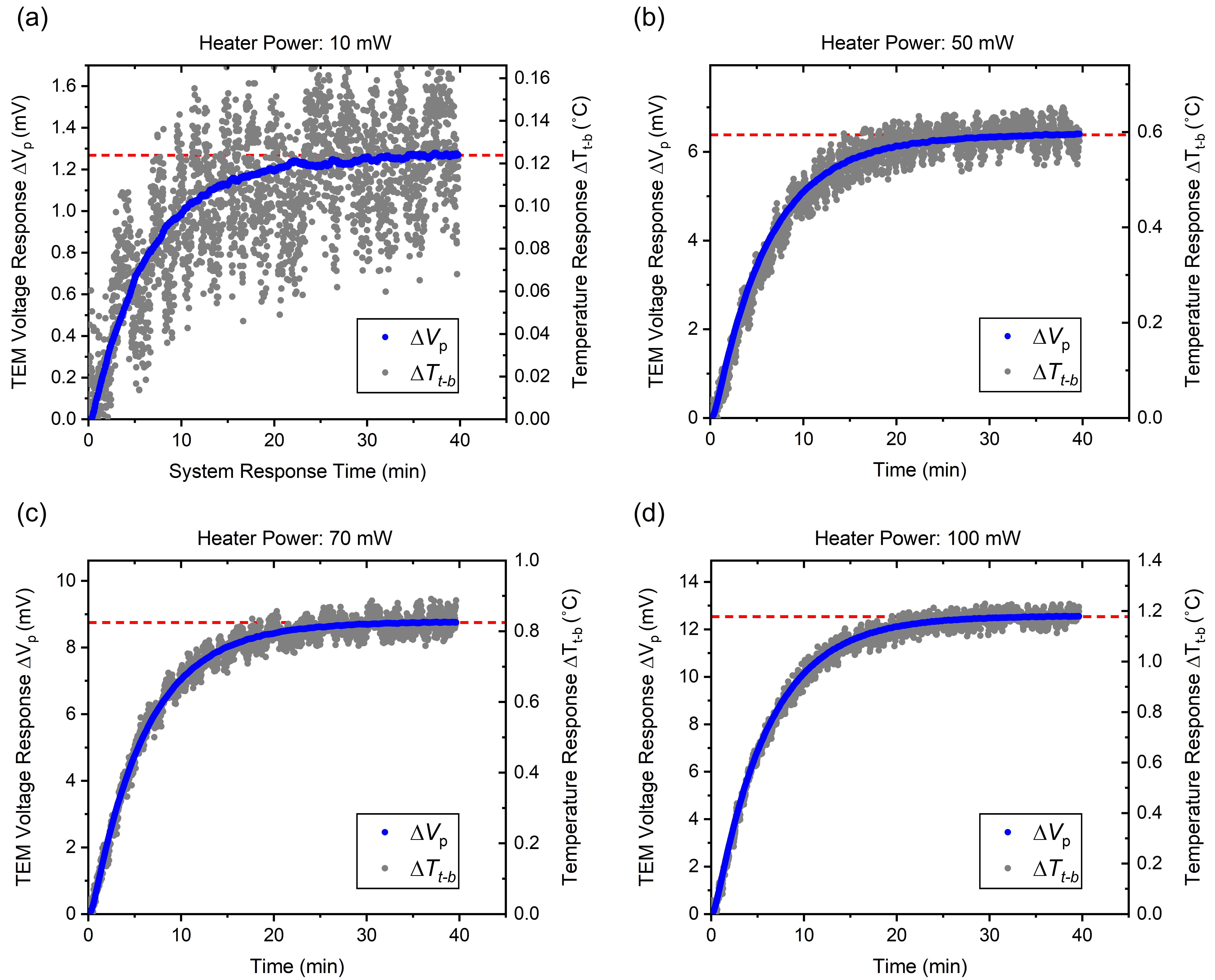}
\caption{Pasive-mode TEM responses when applying (a) 10 mW (b) 50 mW (c) 70 mW (d) 100 mW power to the heat source. The blue curve indicates the measured TEM voltage signal while the gray dots indicate the temperature difference of the TEM, both as a function of time. The time constant of our setup was found to be $\tau_{a} = 6.5905 \ \mathrm{min}$ $\tau_{b} = 6.3702 \ \mathrm{min}$ $\tau_{c} = 6.2837 \ \mathrm{min}$, and $\tau_{d} = 6.19125 \ \mathrm{min}$, respectively. The time constants measured here are consistent for each heater power and is  governed by the system response of the experimental setup. Because the voltage and temperature response follow the same trend, we can conclude that the voltage response is directly proportional to the temperature difference and heat flow through the device.}
\label{passive response}
\end{figure*}

To demonstrate the dynamic response of the TEM heat fluxmeter, we characterized the transient thermal behavior of the entire calibration setup in passive mode by simultaneously measuring the TEM voltage and temperatures at both surfaces of the TEM. Figure \ref{passive response} shows the described transient thermal responses for several different heater powers (i.e., 10 mW, 50 mW, 70 mW, and 100 mW). Excellent agreements between the TEM voltage change ($\Delta V_\mathrm{p}$) and the temperature difference between the top and the bottom of the TEM ($\Delta T_\mathrm{t-b} = T_\mathrm{t} - T_\mathrm{b}$) are observed, clearly demonstrating that the voltage signal in passive mode is linearly proportional to the temperature difference of the TEM (i.e., $V_\mathrm{p} \propto \alpha_\mathrm{eff} \Delta T_\mathrm{t-b}$). However, the TEM voltage is much more precise than the differential temperature measurement in tracking the transient thermal behavior. The uncertainty of $\Delta T_\mathrm{t-b}$ measurement is estimated to be $\pm 0.045^\circ$C at the heater power of 100 mW, which is about 28 times worse than the uncertainty of the TEM voltage when it is converted to temperature by using the effective Seebeck coefficient (i.e., $U_T = U_V/\alpha_\mathrm{eff}=\pm 0.0016^\circ$C). Figure \ref{passive response} also shows that the TEM heat fluxmeter is fast enough to track the relatively slow transient thermal behavior of the experimental setup. The time constant of the experimental setup is $\tau_\mathrm{sys} = 6.36 \pm 0.27$ min as determined by the average time constant from the obtained results for different heater powers. 

\begin{figure*}
\includegraphics[width=\textwidth,height=\textheight,keepaspectratio]{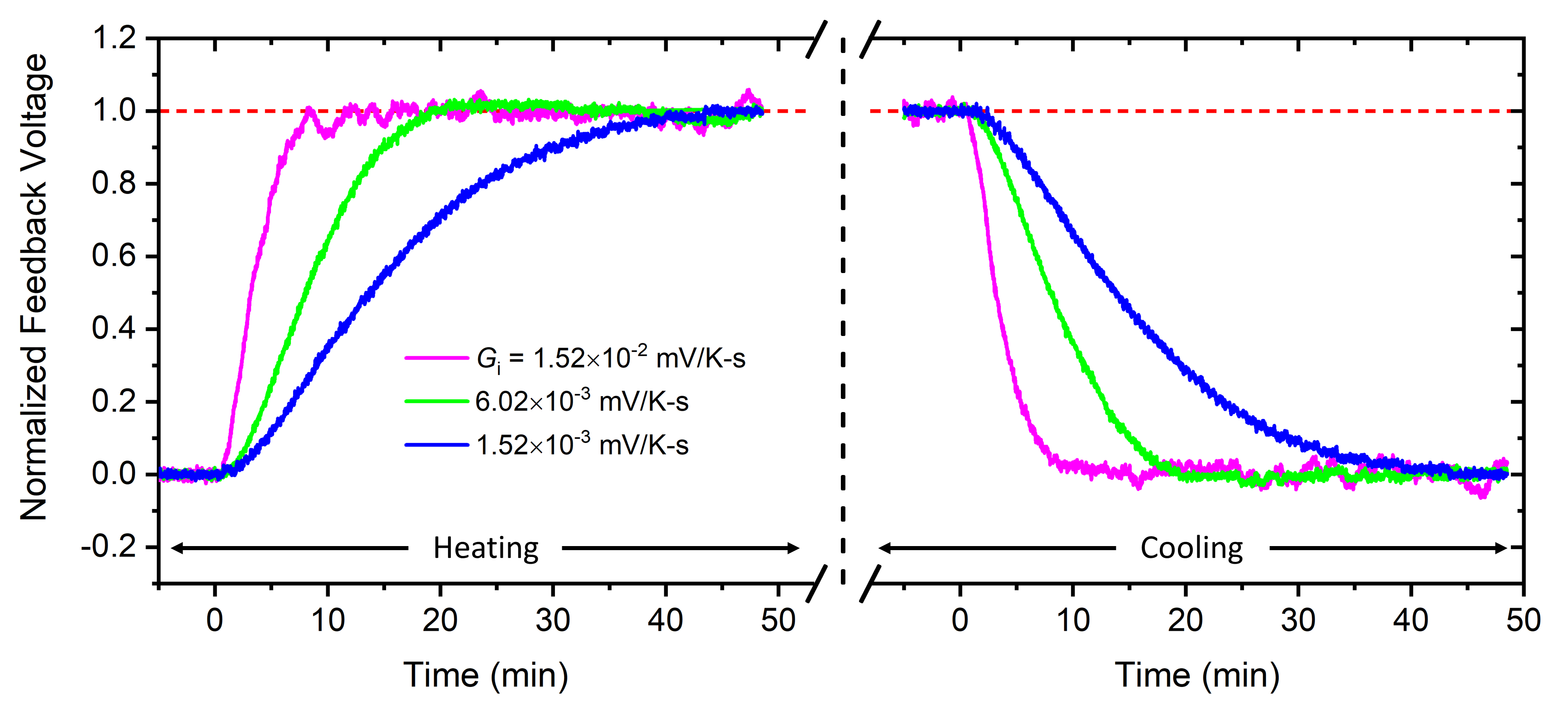}
\caption{The effect of the integral gain $G_{\mathrm{i}}$ on the TEM response. The top TEM temperature was held at $18 \ ^\circ$C, while the integral gain was adjusted. At high gain, the applied TEM voltage signal starts to become unstable, while a low gain has a slow signal response. With proper gain tuning, the response time can be optimized with a stable control.}
\label{active response}
\end{figure*}

In addition to having a higher sensitivity than passive mode, another advantage of active mode can be found in manipulating a transient thermal response of a system using a feedback controller. Figure \ref{active response} shows the TEM voltage applied to maintain $T_\mathrm{t} = 18 ^\circ$C for the heater power of 100 mW. For the present work, the proportional and derivative gains are fixed to $G_\mathrm{P} = 6.816 \times 10^{-1} ~\mathrm{mV/K}$ and $G_\mathrm{D} = 1 \times 10^{-1} ~\mathrm{mV\cdot s/K}$, respectively, while the integral gain $G_\mathrm{I}$ is varied. The integral gain of the feedback controller affects the system's response time to reach the the steady state. When the settling time is defined as time it takes for $V_\mathrm{a}(t)$ to reach within the noise level around a steady state, Fig. \ref{active response} demonstrates that the settling time of the experimental setup for heating can be reduced from $\sim$45 min to $\sim$15 min when increasing the integration gain by ten times from $G_\mathrm{I} = 1.52 \times 10^{-3} $ to $1.52 \times 10^{-2} \mathrm{mV/K\cdot s}$. For comparison, the settling time for passive mode with 100-mW heating is estimated to be $\sim$30 min from Fig. \ref{passive response}(d). A similar result is obtained for cooling although the cooling case takes a slightly longer settling time.   
However, when the integral gain is set too high, the voltage output becomes unstable as observed for $G_\mathrm{I} = 1.52 \times 10^{-2} \mathrm{mV/K\cdot s}$. 
It should be noted that only the effect of the integral gain is discussed here, but we anticipate a similar result for adjustments of the proportional gain, albeit, with a higher sensitivity. The obtained result suggests that with proper gain optimization, active mode can allow quick heat flux measurement by pushing a system to a steady state in a fast, yet, stable manner.

\subsection{Dual TEM Scheme for Better Precision Measurement}
\begin{figure*}
\includegraphics[width=\textwidth,height=\textheight,keepaspectratio]{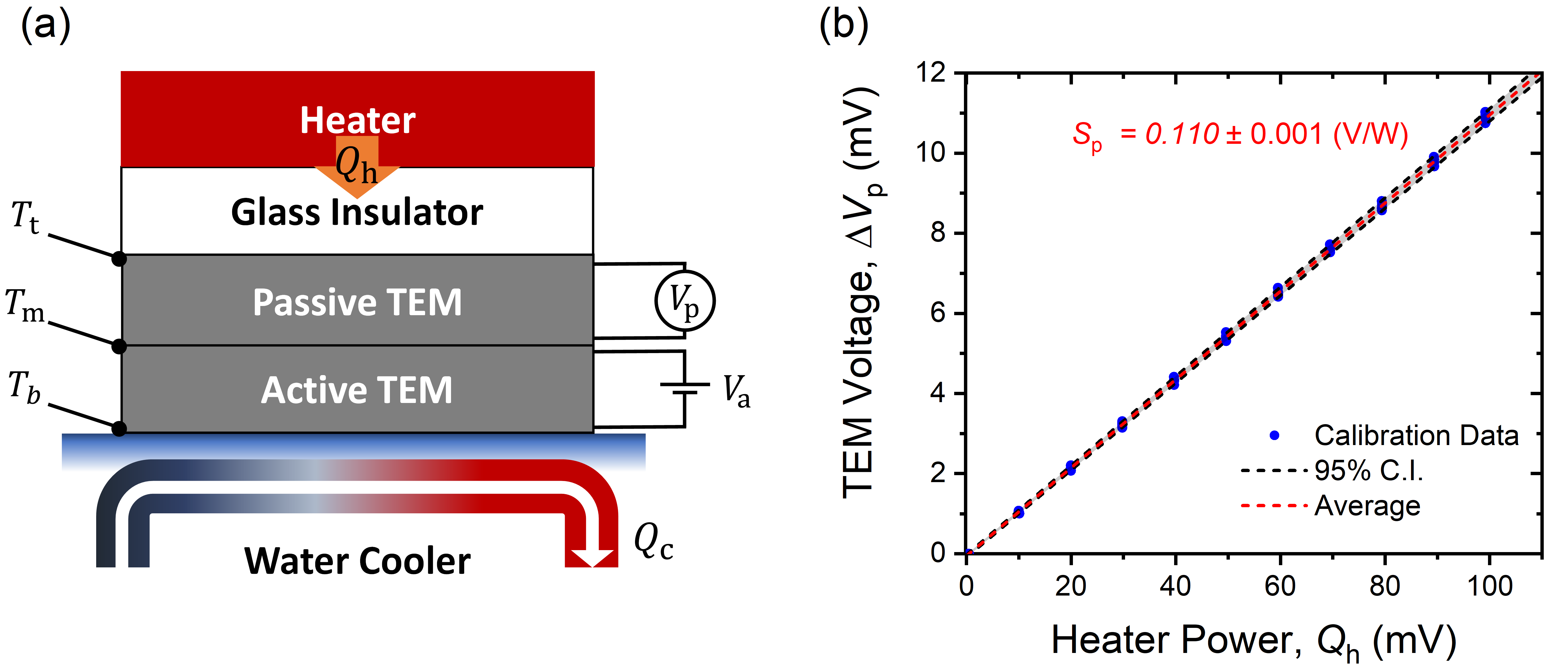}
\caption{(a) Dual TEM experimental setup by combining an active- and passive-mode TEM operation. The active TEM is used to maintain constant bottom temperature of the passive TEM, $T_{\mathrm{b}}$. (b) Shows the new calibration data in dual-stack mode. Compared to Figure \ref{calibration}(a) the percent uncertainty of the sensitivity in our calibration set is reduced from $\pm 3.5 \%$ to $\pm 0.9 \%$}
\label{dual TEM calibration}
\end{figure*}

As shown in Fig. \ref{calibration}, the uncertainty of the TEM sensitivity is determined to be $\pm 0.004$ V/W for passive mode and $\pm 0.020$ V/W for active mode when single TEM is used for heat flux measurement. Such uncertainty gives rise to larger inaccuracy at higher heat flow rates. Since the TEM essentially measures the temperature difference between the top and the bottom surfaces, we suspect that the slope uncertainty is mainly due to unstable $T_\mathrm{b}$ although the water chiller is used as a heat sink. 
As a potential approach to reduce the uncertainty of the TEM sensitivity, we propose a dual TEM scheme by stacking two TEMs on top of each other as illustrated in Fig. \ref{dual TEM calibration}(a). In the dual TEM scheme, the top TEM device is operated in passive mode as a primary heat fluxmeter while the bottom TEM is actively controlled to maintain a constant bottom temperature of the passive TEM device, $T_\mathrm{m}$.  
The steady-state calibration procedure is the same as discussed in the previous section, and the results are shown in Fig. \ref{dual TEM calibration}(b). When compared to the single TEM scheme in passive mode, the dual TEM scheme reduces the uncertainty of the sensitivity from $\pm 0.004$ V/W to $\pm 0.001$ V/W (or from $\pm 3.5\%$ to $\pm 0.9\%$). The obtained result confirms our hypothesis that the bottom temperature should be stable to achieve a high precision of the heat flux measurement. This requirement is particularly crucial for high heat flow rate measurements. It should be noted that the measurement uncertainty could be suppressed as well when the top TEM is operated in active mode in the dual TEM scheme. As discussed earlier, the active-mode operation has a couple of advantages over passive mode, such as a large voltage signal for a given heat flow rate and reducing the response time of a thermal system. However, actively controlling both TEMs requires the simultaneous operation of two PID controllers, resulting in a more complicated measurement setup. We believe that the dual TEM scheme in passive mode is the most desirable approach to precisely measure a heat flux using commercial TEMs with relatively simple experimental configuration. 
\section{Conclusion}
The present work investigates the use a commercial thermoelectric module as a heat fluxmeter. We have calibrated a TEM device in passive and active modes to find a linear relation between the heat flow rate and the TEM voltage signal for both cases. The time constant of the used TEM in size of 15$\times$15 mm$^2$ is measured to be around 10 sec, which is fast enough to measure the transient heat transfer of a thermal system as far as its thermal response is slower than the TEM. Furthermore, the active-mode operation can reduce the systematic response time required to reach a steady state by optimizing the feedback gains of the implemented controller. In order to further improve the measurement precision, we propose a dual TEM scheme as an effective way that stabilizes the bottom temperature of the top TEM (the primary heat fluxmeter) by actively controlling the bottom TEM. The steady-state and dynamic calibration results presented in this work provide a cost-effective way to accurately measure a heat flux in various thermal environments.

\begin{acknowledgments}
This work has been supported by the National Science Foundation (Grants No. CBET-1403072 and No. ECCS-1611320). A.N.M.T.E. acknowledges financial supports from the University of Utah Graduate Research Fellowship. 
\end{acknowledgments}

\nocite{*}
\bibliography{References}

\clearpage

\clearpage

\clearpage

\clearpage

\clearpage

\end{document}